\documentstyle[12pt]{article}

\def\be{\begin{eqnarray}}
\def\ee{\end{eqnarray}}
\def\eq{\label}

\makeatletter
  
  \@addtoreset{equation}{section}
\makeatother

\def\abstract#1{\vskip 7mm 
	\begin{center}{\large Abstract}\par \bigskip
		\begin{minipage}[c]{12cm}
			\small #1
		\end{minipage}
	\end{center}
}
\def\title#1{\begin{center}{\Large\bf #1}\end{center}}
\def\author#1{\vskip 5mm \begin{center}{#1}\end{center}}
\def\address#1{\begin{center}{\it #1}\end{center}}

\newcommand{\bfr}{\begin{flushright}}
\newcommand{\efr}{\end{flushright}}

\begin{document}

\vspace*{-2cm}
\bfr{}\efr\vspace{-9mm}
\bfr{}\efr
\vspace{0.5cm}

\title{Moyal Quantization on Fuzzy Sphere}
\author{Takao KOIKAWA\footnote{E-mail: koikawa@nbilx02.nbi.dk \\On leave absence from Otsuma Women's University}}
\vspace{1cm}
\address{
        The Niels Bohr Institute,\\
         Blegdamsvej 17, 2100 Copenhagen, \\
         Denmark
}
\vspace{5cm}

We study the quantization of compact space on the basis of the Moyal quantization. We first construct the $su(2)$ algebra that are the functions of canonical coordinates $a$ and $a^*$. We make use of them to define the adjoint operators, which is used to define the fuzzy sphere and constitute the algebra. We show that the vacuum is constructed as the powers of $a^*$, in contrast to the flat case where the vacuum is defined by the exponential function of $a$ and $a^*$. We present how the analogy of the creation operator acting on the vacuum is obtaied. The construction does not resort to the ordinary creation and annihilation operators.

\newpage
\setcounter{page}{2}



\section{Introduction}

There has been great interest in the noncommtative spacetime.\cite{rf:1,rf:2} The development of the string theory supports the noncommutative spacetime which is the theory of gravitation at Planch scale. The deformation theory also deals with the noncommutative phasespace variables,\cite{rf:3} and it is based on the abstract star product satisfying some axioms and on a deformation of the symplectic structure of the phasespace in the classical theory. This is formulated without a Hilbert space and the commutation relation using the star product becomes the Poisson bracket as $\hbar \to 0$. This is originated from another way of quantization called the Moyal quantization.\cite{rf:4} The formulation is based on only functions which play the role of both the operators and the states. One of the advantages of using functions instead of using operators is that the geometrical implication is clear in contrast to the operator formulation. Although the Moyal quantization is another way of constructing the quantum mechanics, there exist a relation called the Wigner-Weyl(WW hereafter) correspondence \cite{rf:5} that connects operators with functions called the Weyl symbols. 

There appears noncommutativity in various fields in physics, we should note that the flat coordinates are assumed and the phasespace variables satisfy the canonical commutation relation in the naive Moyal quantization. It may be possible to modify the Moyal quantization in such a way that it can be applied to the curved spacetime, but this is a difficult problem.\cite{rf:6} When there are some second class constraints, it is well known that the Poisson bracket should be taken over by the Dirac bracket for defining the quantization. As a result, the phasespace variables do not satisfy the canonical commutation relation. The application of the Moyal quantization to the constrained system is a challenging subject. Since the canonical commutation relations do not exist any more in such a system, it is different from the naive Moyal quantization so that we call it the function formulation here. In this case, there can not be one-to-one correspondence between the operator and the Weyl symbol in general. 

One of the purposes of investigating noncommutative spacetime is to find some clue in constructing the framework of describing quantum gravity. The lack of the one-to-one correspondence implies that the operator formulation and the function formulation are not connected directly, and so we are compelled to choose whether we start with either the operators or the functions. In this paper, we discuss the fuzzy sphere,\cite{rf:7,rf:8} (see also \cite{rf:9} and the references therein) that is a toy model of the general quantization of the curved space, by using only functions and the star product among them without reference to operators. We adopt the function formulation because of the geometrical reason mentioned above. We also elucidate the relation between the ``states" in the function formulation and the spherical harmonics.    

In the following section, we recapitulate the fuzzy sphere construction that is based upon operators. In section 3, we summarize how we can obtain the spectrum of the one-dimensional harmonic oscillator. The states are to be constructed both algebraically and analytically. The former corresponds to the 2nd quantization while the latter to the 1st quantization. We illustrate the one-dimensional harmonic oscillator in order to compare the infinite dimensional states with the finite dimensional Hilbert states of the fuzzy sphere. In section 4, we construct the fuzzy sphere without resort to the operators. In the construction, no operator appears. The construction is based on canonical variables where the Moyal bracket is used to define the commutation relation(CR hereafter).  We use them to construct the $su(2)$ generators. We then discuss the fuzzy sphere by the adjoint operators acting on the space expressed in terms of the canonical variables. We show how the finite dimensionality appears by defining the vacuum and the creation operators acting on it. The equation among the adjoint operators is regarded as the definition of a fuzzy sphere. The last section is devoted to the summary and discussion.   
  



\section{Operator formulation of fuzzy sphere}

The fuzzy sphere is defined by quantizing the algebra on the 2-sphere. The algebra consists of operators acting on $N+1$ dimensional Hilbert space. Therefore we can regard the function algebra as that of the complex matrices $M_{N+1}(C)$. The construction of the algebra shown in this section is based on the operator formalism.

We introduce 2 pairs of creation-annihilation operators $\hat a^{\dagger}_i$ and $\hat a_i$(i=1,2) which constitute $su(2)$ fundamental representation. They satisfy
\be
\left[\hat a_i,\hat a^{\dagger}_j \right]_{op}=\delta_{ij},
\ee
where $\left[\hat o_1,\hat o_2\right]_{op}=\hat o_1 \hat o_2-\hat o_2\hat o_1$ is the commutation relation between the operators.
We define the $N+1$ dimensional Hilbert space ${\cal F}_N$ spanned by
\be
\left|k \right>=\frac{1}{\sqrt{k!(N-k)!}}\left(\hat a^{\dagger}_1\right)^k\left(\hat a^{\dagger}_2\right)^{N-k}\left|0 \right>,
\ee
where $\left|0 \right>$ is the vacuum annihilated by either $\hat a_1$ or $\hat a_2$. These states are characterized by
\be
{\hat N}\left|k \right >=N\left|k\right>,
\ee
where $\hat N$ is the number operator defined by $\hat N=\hat a^{\dagger}_i \hat a_i$. The algebra ${\cal A}_N$ acting on ${\cal F}_N$ consists of operators commutable with the number operator, and its generators are given by
\be
\hat x_i=\frac{\alpha}{2}\sigma_i^{mn}\hat a^{\dagger}_m \hat a_n,
\label{coordinates}
\ee
where $\sigma_i$ are the Pauli matrices and $\alpha$ is some constant. We can show that the generators satisfy the $su(2)$ relations
\be
\left[{\hat x}_i,{\hat x}_j\right]_{op} &=& i \alpha \epsilon_{ijk}{\hat x}_k.
\label{su2}
\ee
When we impose the constraint on these $x_i$ by
\be
\hat x_i \hat x_i=\frac{\alpha^2}{4}{\hat N}({\hat N}+2)=r^2,
\ee
these generators can be regarded as the noncommutative coordinates of the fuzzy sphere. This corresponds to fixing the angular momentum to be constant and so the remaining degree of freedom is the magnetic quantum number. Here the constant $r$ is interpreted as the radius of the fuzzy sphere. We can define the derivatives or the Killing vector fields $\hat L_i$ of ${\cal A}_N$ by the adjoint action of $x_i$ by
\be
\hat L_i {\hat x}=\frac{1}{\alpha}ad_{\hat x_i}{\hat x}=\frac{1}{\alpha}\left[\hat x_i,\hat x\right]_{op},
\ee
where $\hat x \in {\cal A}_N$. Then we can show that these $\hat L_i$ also satisfy $su(2)$ algebra
\be
\left[{\hat L}_i,{\hat x}_j\right]_{op} &=& i \epsilon_{ijk}{\hat x}_k,\\
\left[{\hat L}_i,{\hat L}_j\right]_{op} &=& i \epsilon_{ijk}{\hat L}_k.
\ee

These show how the fuzzy sphere is constructed by using operators. The role of $\hat x_i$ is twofold. One is the operators acting on the Hilbert space ${\cal F}_N$, and the other is the ``coordinates" of the fuzzy sphere. The constraint induces the introduction of the cutoff $N$ of the angular momentum, and the limiting the spherical harmonics with the angular momentum less than the maximal angular momentum $N$. Note that the commutative limit of eq.(\ref{su2}) is obtained by letting $\alpha$ introduced in eq.(\ref{coordinates}) go to zero, which turns out to take the infinitely large $N$ limit.

As far as we use the operator formalism, the relation between the states, which is based on the vacuum $\left|0\right >$, and the spherical harmonics is not clear, since the cutoff to the angular momentum is now set. We introduce the constant $\alpha$ in eq(\ref{coordinates}) and it plays the role of the structure constant in eq.(\ref{su2}). But its origin is not clear. These motivate the present work. One of the advantages of using the Moyal quantization method is that there appear only functions that play the role of both the states and the operators acting on them.




\section{Moyal quantization of one-dimensional oscillator}

In this section we recapitulate how the eigenvalue problem of the one-dimensional harmonic oscillator is formulated by the Moyal quantization method where both the operators and the states are expressed in terms of functions explicitly. The purpose of this section is to compare the result with case of the finite dimensional Hilbert space discussed in the following section.   

We define the 2-dimensional coordinates $a$ and $a^*$ by
\be
a=\frac{1}{\sqrt 2}(x_1+ix_2),\eq{defofa}
\\
a^*=\frac{1}{\sqrt 2}(x_1-ix_2).\eq{defofastar}
\ee
The star product of two functions $f(a,a^*)$ and $g(a,a^*)$ is defined by
\be
f(a,a^*)\star g(a,a^*)={\rm e}^{\frac{\kappa}{2} 
\Big(\frac{\partial}{\partial a}\frac{\partial}{\partial a'^*}-
\frac{\partial}{\partial a^*}\frac{\partial}{\partial a'}\Big)}f(a,a^*)g(a',a'^*)|_{a=a',a^*=a'^*}.
\eq{eq:starproduct}
\ee
The Moyal bracket of these functions is defined by
\be
\left[f(a,a^*), g(a,a^*)\right] = f(a,a^*)\star g(a,a^*) - g(a,a^*)\star f(a,a^*).
\ee
We discuss the one-dimensional harmonic oscillator whose Hamiltonian is given by
\be
H=\frac{1}{2}\left(x_1^2+x_2^2\right),
\ee
which is to be expressed by using $a$ and $a^*$ as
\be
H=a^* \star a+\frac{\hbar}{2}.
\ee
Here we have used the following Moyal bracket by setting $\kappa=\hbar$
\be
\left[a, a^*\right]=\hbar. 
\eq{canonical}
\ee

The stargenfunctions $w_{nm}$ to the stargen-value equations \cite{rf:10,rf:11}
\be
H\star w_{nm}=E_nw_{nm},\quad w_{nm}\star H =E_m w_{nm},
\ee
with $E_n=(n+1/2)\hbar$ are obtained algebraically by multiplying $a^*$s from the left and $a$s from the right on the vacuum by using the starproduct
\be
w_{nm} \sim (a^* \star)^n w_{00}(\star a)^m, n=0,1,\cdots,~m=0,1,\cdots.
\ee
Here $w_{00}$ is the vacuum given by
\be
w_{00}=2e^{-\frac{2H}{\hbar}},
\ee
which satisfies
\be
a\star w_{00}=w_{00}\star a^*=0.
\ee
Since only functions appear in the stargenvalue problem, 
both the ``operator" and  the stargenfunctions $w_{nm}$ are to be expressed in terms of functions explicitly. They are given by using the the Laguerre polynomials by \cite{rf:6}
\be
w_{nm} \propto  a^{m-n}e^{-\frac{2H}{\hbar}}L_n^{(m-n)}(z).
\ee

The stargenvalue functions constitute the infinite dimesinal vector fields. In the next section, we shall discuss the finite dimensional Hilbert space constructed on a basis of the vacuum function.




\section{Function formulation of fuzzy sphere}

In the previous section we use the notation $x_1$ and $x_2$ to denote the phasespace variables which are a coordinate and a momentum satisfying the canonical relation. Here we consider the 3-dimensional Euclidean space whose coordinates are given by $x_i$, (i=1,2,3). We then define a 2-sphere defined by $x_ix_i=1$, by which we can use $x_1$ and $x_2$ as independent variables on the 2-sphere. Instead of $x_1$ and $x_2$, we use $a$ and $a^*$ defined in (\ref{defofa}) and (\ref{defofastar}), which satisfy the canonical commutation relation $[a,a^*]=\kappa$. We use $\kappa$ as a parameter appearing in the definition of the star product instead of using $\hbar$ as in the preceding section. The functions on the 2-sphere is written as $f=f(a, a^*)$.

In this section we discuss the finite dimensional Hilbert space expressed in terms of functions. 
\be
s_3&=&-\frac{1}{2}aa^*,\\
s_+&=&\frac{1}{2}a^2,\\
s_-&=&-\frac{1}{2}a^{*2}.
\ee
Then we can show that they satisfy the $su(2)$ relation
\be
\left[s_3,s_{\pm}\right]&=&\pm \kappa s_{\pm},\\
\left[s_+,s_-\right]&=&2 \kappa s_3.
\ee
We next define the derivatives or the adjoint operators $\hat l_i$ acting on $f=f(a,a^*)$ by
\be
ad_{s_i}f=\left[s_i,f\right]=\hat l_i f,~i=3,\pm.
\ee
They are explicitly obtained as
\be
{\hat l_3}&=&\frac{\kappa}{2}\left(a\frac{\partial}{\partial a}-a^*\frac{\partial}{\partial a^*}\right),\\
{\hat l_+}&=&\kappa a\frac{\partial}{\partial a^*},\\
{\hat l_-}&=&\kappa a^*\frac{\partial}{\partial a}.
\ee
These operators also satisfy the $su(2)$ relation
\be
\left[\hat l_3,\hat l_{\pm}\right]_{op}&=&\pm \kappa \hat l_{\pm},
\eq{opcr1}
\\
\left[\hat l_+,\hat l_-\right]_{op}&=&2 \kappa \hat l_3.
\eq{opcr2}
\ee
We define $\hat l^2$ by
\be
\hat l^2=\hat l_3^2+\frac{1}{2}\left(\hat l_+\hat l_-+\hat l_-\hat l_+\right),
\eq{fuzzysphere}
\ee
which is to be expressed as
\be
\hat l^2=\frac{{\kappa}^2}{4}{\hat {\cal N}}\left({\hat {\cal N}}+2\right),
\ee
where 
\be
{\hat {\cal N}}=a^*\frac{\partial}{\partial a^*}+a\frac{\partial}{\partial a}.
\ee
We can show that
\be
\left[\hat l^2,\hat l_i \right]_{op}=0,~i=3,\pm.
\ee
This means that we may diagonalize $\hat l^2$ and $\hat l_3$ simultaneously. Let us compute the vectors that diagonalize these operators.
We introduce $N+1$ functions $v_k~(k=0,1,2,\cdots,N)$ by
\be
v_k=\hat l_+^k a^{*N}=\left[s_+, \left[s_+, \cdots \left[s_+,a^N  \right] \right] \cdots \right] \sim a^{k}a^{*(N-k)},
\eq{vkdef}
\ee
which are alternatively given by
\be
v_k \sim \hat l_-^{N-k}a^N.
\ee

Then, since $v_k$ is an eigenstate of $\hat{\cal N}$ with the eigenvalue $N$, we can show that it is the eigenstate of $\hat l^2$: 
\be
\hat l^2v_k&=&\frac{{\kappa}^2}{4}N\left(N+2\right)v_k,
\eq{totalam2}
\ee
and also the eigenstate of $\hat l_3$:
\be
\hat l_3v_k&=&\frac{\kappa}{2} (2k-N)v_k,
\eq{eigenvalue_eqs}
\ee
which is obtained by use of the explicit form of $v_k$ or by use the operator commutation relations given in Eqs.(\ref{opcr2}). By using Eq.(\ref{opcr1}), we can also show that operators $\hat l_+$ and $\hat l_-$ raise and lower the eigenvalue of $\hat l_3$ by $\kappa$, respectively
\be
\hat l_3 \hat l_{\pm}v_k=\frac{\kappa}{2}\left(2k-N\pm2 \right)l_{\pm}v_k.
\ee

We note that there exist an upper and lower bound in these series, 
\be
(\hat l_+)^{N+1}v_0&=&0,\quad \hat l_-v_0=0,
\eq{creation}
\ee
and
\be
(\hat l_-)^{N+1} v_N&=&0, \quad \hat l_+ v_N=0.
\eq{annihilation}
\ee
This shows that an arbitrarily large number of multiplications of $\hat l_{\pm}$ on $v_k$ have zero eigenvalues, and so only finite number of multiplications could be meaningful. If we call $\hat l_+$ a creation operator, it may be possible to call $v_0$ a vacuum. However, in contrast to the vacuum discussed in the preceding section, the series obtained by successively operating $\hat l_+$ on $v_0$ is truncated at finite operations. This shows how we get the finite dimensional vector space.

We next discuss the orthogonality condition of these $v_k$. Since the coordinates $x_1$ and $x_2$ are the coordinates on the 2-sphere, they are to be parametrized by the spherical coordinates $\theta$ and $\phi$ as
\be
x_1&=&\sin \theta \cos \phi,\\
x_2&=&\sin \theta \sin \phi.
\ee
Then $a$ and $a^*$ are expressed as
\be
a&=&\frac{1}{\sqrt 2} \sin \theta e^{i \phi},
\eq{abyspherical}
\\
a^*&=&\frac{1}{\sqrt 2} \sin \theta e^{-i \phi}.
\eq{astarbyspherical}
\ee

We next define the inner product of two functions on the 2-sphere which are the functions of the spherical coordinates. Let $f=f(\theta ,\phi)$ and $g=g(\theta ,\phi)$ be such functions, and then we define the inner product of them by
\be
(f,g)=\int f^*gd\Omega,
\ee
where
\be
d\Omega=\sin \theta d\theta d\phi.
\ee
We can now show that the series of functions defined in (\ref{vkdef}) are mutually orthogonal. Noting that
\be
\int \left(a^ka^{*(N-k)}\right)^*a^la^{*(N-l)} d\Omega=\cases{
\frac{4 \pi N!}{(2N+1)!!},\quad {\rm for}\quad k=l,\cr
0,\quad {\rm for}\quad k \ne l,
}
\ee
we redefine $v_k$ by
\be
v_k=\sqrt{\frac{(2N+1)!!}{4\pi N!}}a^ka^{*(N-k)}.
\ee
Then we obtain the orthonormal Hilbert space basis satisfying
\be
(v_k,v_l)=\delta_{k,l}.
\ee
This completes the proof that the series $v_k~(k=0,1,\cdots,N)$ constitute the N+1 dimensional vector space.

In order to fix the dimension of the vector space, we impose a constraint
\be
\hat l^2=R^2,
\eq{fuzzysphere}
\ee
where $R$ is some constant.  It must be clear that, by taking Eq.(\ref{totalam2}) into account, this turns out to be fixing the dimension of the vector space. 
To be more exact ${\kappa}^2N(N+2)$ is fixed, therefore we find that the commutative limit $\kappa \to 0$ corresponds to $N \to \infty$ limit.

Now that the algebra $\cal O$ consisting of operators $\hat l_i$ acts on the $N+1$ dimensional vector space and its adjoint vector space, we can regard it to be equivalent to the $(N+1)\times(N+1)$ matrices. Those operators $\hat l_i$, which are first introduced as ``operators" acting on functions of $a$ and $a^*$, can be viewed in a different way. 
Another aspect of noncommutative $\hat l_i$ arises from how we view Eq.(\ref{fuzzysphere}). We regard the equation as the definition of the fuzzy sphere. We might call it so by implementing them with the derivative of $\hat L \in {\cal O}$ along $\hat l_i$ defined by
\be
\frac{1}{\kappa}{\widehat {ad}}_{\hat l_i} \hat L=\frac{1}{\kappa}\left[\hat l_i,\hat L\right]_{op}=\hat K_i \hat L.
\ee
We can show that they also satisfy the $su(2)$ relation as before,
\be
\left[\hat K_3,\hat K_{\pm}\right]_{op}&=&\pm  \hat K_{\pm},\\
\left[\hat K_+,\hat K_-\right]_{op}&=&2 \hat K_3.
\eq{kvcr2}
\ee

In the remaining of this section, we express $v_k$ in terms of the spherical coordinates and investigate their relation to the spherical harmonics. The operators and their eigenestates in the Moyal quantization formalism are both constructed by functions, as was exemplified in preceding section. We restrict $N$ to be even numbers, $N=2l$ where $l$ is the nonnegative integer. The restriction comes from the fact that the spherical harmonics are the representation of $so(3)$ rather than $su(2)$. We can express the eigenstates $v_k$ in terms of the spherical coordinates by use of Eqs.(\ref{abyspherical}) and (\ref{astarbyspherical}):
\be
v_k \sim a^ka^{*(2l-k)}=\left(\frac{\sin\theta}{\sqrt2} \right)^{2l} {\rm e}^{2i(k-l)\phi}.
\ee
The adjoint operators $\hat l_i$ are also expressed by using the spherical coordinates. Especially, $\hat l_3$ becomes simple in the spherical coordinates
\be
\hat l_3=-i\frac{\kappa}{2}\frac{\partial}{\partial \phi}.
\eq{l3}
\ee
We reproduce the result in Eq.(\ref{eigenvalue_eqs}) also in the spherical coordinate expression, and find that the ``creation" and ``annihilation" operators correspond to increase and decrease of the angle in the $\phi$ direction.

When $k$ takes special values in the above expression, they are explicitly connected with the spherical harmonics,
\be
v_0 &\sim & a^{*2l}=\left(\frac{\sin\theta}{\sqrt2} \right)^{2l} {\rm e}^{-2il\phi} \propto \left(Y_l^{-l}\right)^2,\\
v_{2l} &\sim & a^{2l}=\left(\frac{\sin\theta}{\sqrt2} \right)^{2l} {\rm e}^{2il\phi} \propto \left(Y_l^l\right)^2.
\ee

Since the dimension of the Hilbert space is finite, the integration over the space is taken over all possible states. Let $\hat O$ be an element of $\cal O$, then the average of $\hat O$ can be computed by
\be
\left<\hat O\right>=\frac{\sum_k \left( v_k,\hat O v_k\right)}{\sum_k \left( v_k, v_k\right)}.
\ee




\section{Summary and discussion}
In this paper, we have constructed the fuzzy sphere in terms of the functions of canonical coordinates $a$ and $a^*$. Defining the CR of $a$ and $a^*$ by the Moyal bracket, we first construct three functions of $a$ and $a^*$ that constitute the $su(2)$ algebra. In contrast to the one-dimensional harmonic oscillator case shown in section 3, they are not used as ``operators" acting on the state. They are used to define the adjoint operators acting on the state functions of $a$ and $a^*$. The orthonormal basis of these functions are constructed by repeatedly operating ``creation operator" $\hat l_+$ on the ``vacuum". In order to fix the dimension of the basis, we impose a constraint among the adjoint operators. As a result, in the finite number of operations of the creation operator, it vanishes. This guarantees the finiteness of the Hilbert space. It is possible to interpret the constraint as the definition of fuzzy square. The constant radius implies that the limit to the infinite dimension of the Hilbert space corresponds to the commutative limit of the adjoint operators, and vice versa. The fuzzy square is implemented geometrically by defining the derivative, which is the Lie derivative in the Moyal version.

We can now compare the flat case quantization with the compact case quantization on the basis of the Moyal quantization. The function form of the vacuum reflects the dimensionality of the Hilbert space. The vacuum of the flat case is expressed in terms of the exponential function, on the other hand, the vacuum of the compact case is expressed by using the powers of $a$ and $a^*$, which guarantees the finiteness of the Hilbert space. In the flat case, there exist canonical variables, which enables us to regard them as creation and annihilation ``operators" in the Moyal quantization. However, when we have no such variables, we need to find their equivalences. We find that not the quantities that satisfy the $su(2)$ relations but their adjoint operators, which also satisfy the $su(2)$ relations, include such equivalences. One of them $\hat l_3$ is diagonalized by the states and the remainings are used to change the eigenvalues of the states by $\kappa$. We thus interprete them as the creation and annihilation operators.

The method developed here is quite general, and so it can be applied to other algebra. In such applications we need pairs of $a_i$ and $a^*_i$ satisfying the canonical CRs. One of the advantages of using CRs, which is expressed in terms of the Moyal bracket, is that we use only c-number functions that play the role of operators and the states. Therefore, it is easy to give various equations the geometrical meaning, and so intuitive approach becomes feasible. 

We should note that there does not exist a WW correspondence in general, if the variables do not satisfy the canonical CRs. We can now compare the fuzzy sphere constructions by using the operators shown in section 2 and the functions of $a$ and $a^*$ developed in the previous section. They are not completely in parallel. In defining the vector states, we use only a pair of canonical variables while two pairs of canonical operators are used to define the basis. 
The problem of operator formulation lies in the difficulty to find the corresponding functions isomorphic to operators and states. It is not clear how we can find an equivalence to the vacuum in functions. On the other hand, in the function formulation of the fuzzy sphere, we do not need to find the isomorphism by definition.

\end{document}